\documentstyle[12pt]{article}                    
\newfont{\ffont}{msym10}                        
\newcommand{\beq}{\begin{equation}}             
\newcommand{\eeq}{\end{equation}}               
\newcommand{\bqry}{\begin{eqnarray}}            
\newcommand{\eqry}{\end{eqnarray}}              
\newcommand{\bqryn}{\begin{eqnarray*}}          
\newcommand{\eqryn}{\end{eqnarray*}}            
\newcommand{\NL}{\nonumber \\}                  
\newcommand{\preprint}[1]{\begin{table}[t]      
            \begin{flushright}                  
            \begin{large}{#1}\end{large}        
            \end{flushright}                    
            \end{table}}                        
\newcommand{\PD}[2]                             
    {\frac{\partial^{#2}}{\partial #1^{#2}}}    
\begin{document}
\preprint{LA-UR-98-492}
\title{On the Regge Slopes Intramultiplet Relation}
\author{\\ L. Burakovsky\thanks{E-mail: BURAKOV@T5.LANL.GOV} \ and \  
T. Goldman\thanks{E-mail: GOLDMAN@T5.LANL.GOV} \
\\  \\  Theoretical Division, MS B285 \\  Los Alamos National Laboratory \\ 
Los Alamos, NM 87545, USA \\} 
\date{ }
\maketitle
\begin{abstract}
We show that only additivity of inverse Regge slopes is consistent with both 
the formal chiral limit $m(n)\rightarrow 0$ and the heavy quark limit $M(Q)\gg
M(n),$ where $n=u,d,$ and $m,M$ are current and constituent quark masses, 
respectively. 
\end{abstract}


\bigskip

\section*{Introduction}
It is well known that the hadrons composed of light $(u,d,s)$ quarks populate 
linear Regge trajectories; i.e., the square of the mass of a state with 
orbital momentum $\ell $ is proportional to $\ell :$ $\ell =\alpha ^{'}M^2(
\ell )+a(0),$ where the slope $\alpha ^{'}$ depends weakly on the flavor 
content of the states lying on the corresponding trajectory,
\beq
\alpha ^{'}_{n\bar{n}}\simeq 0.88\;{\rm GeV}^{-2},\;\;\;   
\alpha ^{'}_{s\bar{n}}\simeq 0.84\;{\rm GeV}^{-2},\;\;\;   
\alpha ^{'}_{s\bar{s}}\simeq 0.80\;{\rm GeV}^{-2}.   
\eeq
In contrast, the data on the properties of Regge trajectories of hadrons 
containing heavy quarks are almost nonexistent at the present time, although
it is established \cite{BB} that the slope of the trajectories decreases with
increasing quark mass (as seen in (1)) in the mass region of the lowest
excitations. This is plausibly due to an increasing (with mass) contribution 
of the color Coulomb interaction, leading to a curvature of the trajectory near
the ground state \cite{JN}. However, as the analyses \cite{BB,KS,QR} show, 
in the asymptotic regime of the highest excitations, the trajectories of both 
light and heavy quarkonia are linear and have the same slope $\alpha ^{'}\simeq
0.9$ GeV$^{-2},$ in agreement with natural expectations from the string model. 

Knowledge of Regge trajectories in the scattering region, i.e., at $t<0,$ and 
of the intercepts $a(0)$ and slopes $\alpha ^{'}$ is also useful for many 
non-spectral purposes, for example, in the recombination \cite{rec} and 
fragmentation \cite{fra} models. Therefore, as pointed out in ref. \cite{BB}, 
the slopes and intercepts of the Regge trajectories are the fundamental 
constants of hadron dynamics, perhaps generally more important than the mass 
of any particular state. Thus, not only the derivation of mass relations 
\cite{BGH1,BGH2,BGH3,BGH4} but also the determination of the parameters $a(0)$
and $\alpha ^{'}$ of heavy quarkonia is of great importance, since they afford
opportunities for better understanding of the dynamics of the strong 
interactions in the processes of production of charmed and beauty hadrons at 
high energies, and estimations of their production rates.

If one assumes the (quasi)-linear form of Regge trajectories for hadrons with 
identical $J^{PC}$ quantum numbers (i.e., belonging to a common multiplet), one
will obtain for the states with orbital momentum $\ell $ 
\bqryn
\ell & = & \alpha ^{'}_{i\bar{i}}m^2_{i\bar{i}}\;+a_{i\bar{i}}(0), \\    
\ell & = & \alpha ^{'}_{j\bar{i}}m^2_{j\bar{i}}\;\!+a_{j\bar{i}}(0), \\    
\ell & = & \alpha ^{'}_{j\bar{j}}m^2_{j\bar{j}}+a_{j\bar{j}}(0).    
\eqryn
Further, the following relation among the intercepts exists:
\beq
a_{i\bar{i}}(0)+a_{j\bar{j}}(0)=2a_{j\bar{i}}(0).
\eeq
This relation was first derived for $u(d)$- and $s$-quarks in the 
dual-resonance model \cite{KKY}. It is satisfied in two-dimensional QCD
\cite{BESW}, the dual-analytic model \cite{KMP}, and the quark bremsstrahlung
model \cite{DB}. Also, it saturates inequalities for Regge trajectories 
\cite{IY} which follow from the $s$-channel unitarity condition. Hence, it
may be considered as firmly established and may transcend specific models.
  
With (2), one obtains from the above three relations, the quadratic mass 
formula
\beq
\alpha ^{'}_{i\bar{i}}m^2_{i\bar{i}}+\alpha ^{'}_{j\bar{j}}m^2_{j\bar{j}}=
2\alpha ^{'}_{j\bar{i}}m^2_{j\bar{i}}.
\eeq

In contrast to the relation among the intercepts, Eq. (2), a relation among 
the Regge slopes is not firmly established yet. Two such relations have been 
proposed in the literature,
\beq
\alpha ^{'}_{i\bar{i}}\cdot \alpha ^{'}_{j\bar{j}}=\left( \alpha ^{'}_{j\bar{i
}}\right) ^2,
\eeq
which follows from the factorization of residues of the $t$-channel poles
\cite{first,KY}, and
\beq
\frac{1}{\alpha ^{'}_{i\bar{i}}}+\frac{1}{\alpha ^{'}_{j\bar{j}}}=\frac{2}{
\alpha ^{'}_{j\bar{i}}},
\eeq
based on topological expansion and the $q\bar{q}$-string picture of hadrons
\cite{Kai}. Also, alternative relations have been also proposed which do not 
agree with either (4) or (5); for example,\footnote{In the following, $m(q)$ 
and $M(q)$ stand for current and constituent masses of quark $q,$ 
respectively.}
\beq
\alpha ^{'}_{j\bar{i}}=\frac{\alpha ^{'}}{1+0.2\left( \frac{M(i)+M(j)}{
{\rm GeV}}\right) ^{3/2}}, 
\eeq
where $\alpha ^{'}\cong 0.88$ GeV$^{-2}$ is the standard Regge slope in the
light quark sector. This last was suggested by Filipponi and Srivastava 
\cite{FS}, and implies that the relation between the slopes is 
\beq
\left( \frac{1}{\alpha ^{'}_{i\bar{i}}}-\frac{1}{\alpha ^{'}}\right) ^{2/3}+
\left( \frac{1}{\alpha ^{'}_{j\bar{j}}}-\frac{1}{\alpha ^{'}}\right) ^{2/3}\!\!
=2\left( \frac{1}{\alpha ^{'}_{j\bar{i}}}-\frac{1}{\alpha ^{'}}\right) ^{2/3}.
\eeq
For light quarkonia (and small differences in the $\alpha ^{'}$ values), there
is no essential difference between Eqs. (4) and (5); viz., for $\alpha ^{'}_{j
\bar{i}}=\alpha ^{'}_{i\bar{i}}/(1+x),$ $x\ll 1,$ Eq. (5) gives $\alpha ^{'}_{
j\bar{j}}=\alpha ^{'}_{i\bar{i}}/(1+2x),$ whereas Eq. (4) gives $\alpha ^{'}_{
j\bar{j}}=\alpha ^{'}_{i\bar{i}}/(1+x)^2\approx \alpha ^{'}/(1+2x),$ i.e, 
essentially the same result to order $x^2.$ Eq. (7), however, differs from 
(4),(5) already for small $x:$ for $i=n,$ it gives $\alpha ^{'}_{j\bar{j}}=
\alpha ^{'}_{i\bar{i}}/(1+2^{3/2}\;\!x)\approx \alpha ^{'}_{i\bar{i}}/(1+2.83
\;\!x).$ For heavy quarkonia (and expected large differences from the $\alpha 
^{'}$ values for the light quarkonia) Eqs. (4),(5) are incompatible; e.g., for
$\alpha ^{'}_{j\bar{i}}=\alpha ^{'}_{i\bar{i}}/2,$ Eq. (4) will give $\alpha ^{
'}_{j\bar{j}}=\alpha ^{'}_{i\bar{i}}/4,$ whereas Eq. (5) $\alpha ^{'}_{j\bar{j
}}=\alpha ^{'}_{i\bar{i}}/3.$ Eq. (7) gives, respectively, $\alpha ^{'}_{j\bar{
j}}\approx \alpha ^{'}_{i\bar{i}}/3.83$ in this case. It would be therefore 
useful to establish whether any of these relations is realized in nature. 

In a series of our previous publications \cite{BGH1,BGH2,BGH3,BGH4} we chose 
Eq. (5), since it is much more consistent with (3) than is Eq. (4), when 
tested by using measured quarkonia masses in Eq. (3). By eliminating the 
values of the Regge slopes from Eqs. (3),(5), we derived new (higher power) 
mass relations which hold with high accuracy for all well established meson 
multiplets, and may be reduced to quadratic formulas by fitting the values of 
the slopes. 

Here we wish to compare Eqs. (4),(5), and (7) in both the chiral and heavy 
quark limits. We shall show that, although the three pairs of equations, 
(3),(4), (3),(5) and (3),(7), are consistent in the formal chiral limit $m(n)
\rightarrow 0$ $(n=u$ or $d,$ and we assume SU(2) flavor symmetry: $m(n)\equiv
m(u)=m(d)),$ only one pair of equations, (3),(5), is consistent in the heavy
quark limit $M(Q)\gg M(n),$ thus unambiguously indicating its preferability. 
The same arguments indicate that any other relation among the slopes which 
does not agree with (5) (e.g., Eq. (7)) is inconsistent with the heavy quark 
limit. This confirms the conclusion, drawn before on the basis of meson 
spectroscopy \cite{BGH1,BGH2,BGH3,BGH4}, that it is Eq. (5) that is realized 
in the real world.     

\section*{Formal chiral limit $m(n)\rightarrow 0$}
To consider the formal chiral limit $m(n)\rightarrow 0,$ we use the following
parametrization of the meson masses in terms of the current quark mass, 
discussed in more detail in refs. \cite{Scad,BH}:
\beq
m^2_{n\bar{n}}=2Am(n)+B, 
\eeq
where $A$ and $B$ are constants within a given meson multiplet, but may be
different for different multiplets (note that $B>0$ for non-Goldstone
bosons). As is easily seen, the Regge recurrences of the state (8) have the 
masses $$2Am(n)+B+\frac{1}{\alpha ^{'}_{n\bar{n}}},\;\;2Am(n)+B+\frac{2}{
\alpha ^{'}_{n\bar{n}}},\;\;{\rm etc.}$$

Let us first introduce 
\beq
x_q\equiv \frac{\alpha ^{'}_{q\bar{q}}}{\alpha ^{'}_{n\bar{n}}}\leq 1.
\eeq
Determining now $\alpha ^{'}_{q\bar{n}}/\alpha ^{'}_{n\bar{n}}$ from (4),(5), 
or (7), Eq. (3) may be cast into the form, respectively,
\beq
m^2_{n\bar{n}}+x_q\;\!m^2_{q\bar{q}}=2\sqrt{x_q}\;\!m^2_{q\bar{n}}
\eeq
with Eq. (4), 
\beq
m^2_{n\bar{n}}+x_q\;\!m^2_{q\bar{q}}=\frac{4x_q}{1+x_q}\;\!m^2_{q\bar{n}}
\eeq
with Eq. (5), and 
\beq
m^2_{n\bar{n}}+x_q\;\!m^2_{q\bar{q}}=\frac{2}{1+2^{-3/2}\left( \frac{1}{x_q}-1
\right) }\;\!m^2_{q\bar{n}}
\eeq
with Eq. (7).

Consider, e.g., Eq. (11). As follows from (8),
\beq
m^2_{q\bar{n}}=\frac{1+x_q}{4}\;\!m^2_{q\bar{q}}+\frac{1+x_q}{4x_q}\left( 
2Am(n)+B\right).
\eeq
In the formal limit $m(n)\rightarrow 0,$ $m^2_{n\bar{n}}\rightarrow B,$ and
the masses of its Regge recurrences go in this limit to $$B+\frac{1}{\alpha ^{
'}_{n\bar{n}}},\;\;B+\frac{2}{\alpha ^{'}_{n\bar{n}}},\;\;{\rm etc.,}$$ so 
that all these states again populate Regge trajectory with the initial slope 
$\alpha ^{'}_{n\bar{n}}.$ Thus, the parameter $x_q$ need not change in the 
formal limit $m(n)\rightarrow 0,$ and, as follows from (13),
\bqry
m^2_{q\bar{n}} & \rightarrow  & \frac{1+x_q}{4}\;\!m^2_{q\bar{q}}+\frac{1+x_
q}{4x_q}B  
 \NL
 & = & \frac{1+x_q}{4}\;\!m^2_{q\bar{q}}+\frac{1+x_q}{4x_q}m^2_{n\bar{n}},  
\eqry
which is equivalent to (11). We conclude, therefore, that Eq. (11) holds in the
formal chiral limit $m(n)\rightarrow 0.$ Following a similar procedure, it may
be shown that both Eqs. (10) and (12) also hold in this limit. Thus, this 
limit does not distinguish between the three possible flavor dependent Regge 
slope relations.

\section*{Heavy quark limit $M(Q)\gg M(n)$}
Consider now the heavy quark limit $M(Q)\gg M(n),$ and start with Eqs. (3),(5).
Since the slope decreases with the increasing quark mass, one expects $\alpha
^{'}_{Q\bar{Q}},\alpha ^{'}_{Q\bar{n}}\ll \alpha ^{'}_{n\bar{n}},$ so that one
may neglect $1/\alpha ^{'}_{n\bar{n}}$ in comparison with $1/\alpha ^{'}_{Q
\bar{Q}},$ $1/\alpha ^{'}_{Q\bar{n}}$ in Eq. (5); it therefore takes on the 
form
\beq
\alpha ^{'}_{Q\bar{n}}\simeq 2\alpha ^{'}_{Q\bar{Q}}.
\eeq
Also, the term $\alpha ^{'}_{n\bar{n}}m^2_{n\bar{n}}$ is negligible in 
comparison with $\alpha ^{'}_{Q\bar{Q}}m^2_{Q\bar{Q}},$ $\alpha ^{'}_{Q\bar{
n}}m^2_{Q\bar{n}}$ in Eq. (3) in this limit, for all three cases we are 
discussing. Indeed, in any of these cases, in the heavy quark limit $M(Q)
\gg M(n),$ $\alpha ^{'}$ decreases like $\sim 1/(M(Q))^a,$ $0<a<2.$ This 
follows from, e.g., Eq. (6), and Eqs. (17),(18) below in the remaining two 
cases. Therefore, since both $m^2_{Q\bar{n}}$ and $m^2_{Q\bar{Q}}$ grow like 
$\sim (M(Q))^2,$ it is clear that, as $M(Q)\gg M(n),$\footnote{Also, $\alpha ^{
'}_{n\bar{n}}m^2_{n\bar{n}}$ is fixed at $\stackrel{<}{\sim }O(1$ GeV$^2).$} 
\beq
\alpha ^{'}_{n\bar{n}}m^2_{n\bar{n}}\ll \alpha ^{'}_{Q\bar{n}}m^2_{Q\bar{
n}},\alpha ^{'}_{Q\bar{Q}}m^2_{Q\bar{Q}}. 
\eeq
     
We now show that, as $M(Q)\gg M(n),$ $\alpha ^{'}_{Q\bar{Q}}$ decreases like
$\sim 1/(M(Q))^a,$ $0\!<\!a\!<\!2.$ 

For Eq. (5), this follows from the following form of the parametrization of 
the dependence of the slope on the quark masses, consistent with (5), which 
will be discussed in more detail elsewhere \cite{prep}:
\beq
\alpha ^{'}_{j\bar{i}}=\frac{4}{\pi }\frac{\alpha ^{'}}{1+\sqrt{\alpha ^{'}}
\frac{M(i)+M(j)}{2}},
\eeq
where $\alpha ^{'}=\alpha ^{'}_{n\bar{n}}\cong 0.88$ GeV$^{-2}$ is the standard
Regge slope in the light quark sector. In Table I we present the numerical 
values of the parameter $x_q,$ as defined in (9), $(q=n,s,c,b)$ given by both 
Eq. (11) in which the measured vector and tensor meson masses are used, and 
Eq. (17). One sees that the formula (17) is in excellent agreement with 
experiment.    

\begin{center}
\begin{tabular}{|c|c|c|c|c|} \hline
 $q:$ & n & s & c & b   \\ \hline
 Eq. (11) & 1 & $0.89\pm 0.02$ & $0.50\pm 0.01$ & $0.23\pm 0.01$  \\ \hline
 Eq. (16) & 1.001 & 0.889 & 0.499 & 0.231  \\ \hline
\end{tabular}
\end{center}
{\bf Table I.} Comparison of the numerical values of the parameter $x_q,$ $q=
n,s,c,b,$ given by Eq. (11) in which the measured vector and tensor meson 
masses are used, and Eq. (17) in which the following constituent quark masses 
are used (in GeV) (as extracted from $S$-wave meson spectroscopy):
$M(n)=0.29,$ $M(s)=0.46,$ $M(c)=1.65,$ $M(b)=4.80.$ 
 \\

For Eq. (4), a search for a similar form of the parametrization of the slope
dependence on the quark mass, which would be consistent with both Eq. (4) and 
data, leads to 
\beq
\alpha ^{'}_{j\bar{i}}=\frac{C}{\Big[ M(i)M(j)\Big] ^{a/2}},\;\;\;a\approx 
0.32,\;\;\;C\approx 0.59\;{\rm GeV}^{-1.68},
\eeq
and therefore, $\alpha ^{'}_{j\bar{j}}/\alpha ^{'}_{i\bar{i}}=(M(i)/M(j))^{
0.32}.$  In Table II we present the numerical values of the parameter $x_q,$ 
as defined in (9), $(q=n,s,c,b)$ given by both Eq. (10) in which again the  
measured vector and tensor meson masses are used, and Eq. (18). One sees that 
the formula (18) is in excellent agreement with experiment, as well as Eq. 
(17).

\begin{center}
\begin{tabular}{|c|c|c|c|c|} \hline
 $q:$ & n & s & c & b   \\ \hline
 Eq. (10) & 1 & $0.87\pm 0.02$ & $0.59\pm 0.01$ & $0.39\pm 0.01$  \\ \hline
 Eq. (18) & 0.996 & 0.860 & 0.571 & 0.406  \\ \hline
\end{tabular}
\end{center}
{\bf Table I.} The same as in Table I, for Eqs. (10),(18).
 \\

Although the Regge slope dependence on the quark masses (18), $\alpha ^{'}_{j
\bar{i}}\approx \frac{0.6}{[M(i)M(j)]^{1/6}},$ represents certain academic 
interest, it is not realized in the real world, as well as Eq. (6), as we show
below, nor it is well defined in the limit $M(q)\rightarrow 0,$ in contrast to 
its counterparts (6) and (17). 

With (16), Eq. (3) reduces to
\beq
\alpha ^{'}_{Q\bar{Q}}m^2_{Q\bar{Q}}\simeq 2\alpha ^{'}_{Q\bar{n}}m^2_{Q\bar{
n}}.
\eeq
It then follows from (15),(19) that, independent of the numerical values of 
the slopes,
\beq   
m_{Q\bar{Q}}\simeq 2m_{Q\bar{n}},
\eeq
in agreement with the heavy quark limit $M(Q)\gg M(n).$ Hence, the pair of 
equations (3),(5) is consistent in this limit.

A similar procedure for the remaining two pairs of relations, (3),(4) and 
(3),(7), leads, repsectively, to
\beq
m_{Q\bar{Q}}\simeq \sqrt{\frac{2\alpha ^{'}_{n\bar{n}}}{\alpha ^{'}_{Q\bar{
n}}}}\;m_{Q\bar{n}},
\eeq
and
\beq
m_{Q\bar{Q}}\simeq 2^{5/4}\;\!m_{Q\bar{n}}.
\eeq
Eq. (21) is obtained by neglecting the term $\alpha ^{'}_{n\bar{n}}m^2_{n\bar{
n}}$ in (3), while Eq. (22) is obtained by neglecting both this term in (3) and
$1/\alpha ^{'}_{n\bar{n}}$ in (7). It is seen that Eq. (22) is in clear 
contradiction with the heavy quark limit, as given by (20). Requiring Eq. (21)
to be consistent with (20) leads to the following constraint on the slopes in 
the heavy quark limit:
\beq
\alpha ^{'}_{Q\bar{Q}}\rightarrow \frac{\alpha ^{'}_{n\bar{n}}}{4},\;\;\; 
\alpha ^{'}_{Q\bar{n}}\rightarrow \frac{\alpha ^{'}_{n\bar{n}}}{2}. 
\eeq

Since the pair of equations (3),(7) fails already in the heavy quark limit for 
mesons, we shall not examine it any further. We shall, however, examine the two
remaining pairs in the heavy quark limit for baryons, in order to determine 
further which is preferable, since Eq. (21) may still be consistent with the 
heavy quark limit, provided the validity of Eq. (23). Although the slopes of 
heavy quark trajectories, as extracted from data for this case,  do not 
contradict (23), as seen in Table II, we shall show below that the 
generalization of Eq. (4) to baryons does contradict the corresponding heavy 
quark limit.  
 
\section*{Generalization to baryons}
The above analysis may be easily generalized to baryons. In this case, one has
two pairs of relations \cite{BGH2}, which represent the counterparts of Eq. 
(3),
\bqry
\alpha ^{'}_{nnn}m^2_{nnn}\;+\;\alpha ^{'}_{QQn}m^2_{QQn} & = & 2\alpha ^{'}_{
Qnn}m^2_{Qnn}, \\
\alpha ^{'}_{Qnn}m^2_{Qnn}\;+\;\alpha ^{'}_{QQQ}m^2_{QQQ} & = & 2\alpha ^{'}_{
QQn}m^2_{QQn}, 
\eqry
and Eqs. (4) and (5), respectively:
\bqry
\alpha ^{'}_{nnn}\cdot \alpha ^{'}_{QQn} & = & \left( \alpha ^{'}_{Qnn}\right)
^2, \\
\alpha ^{'}_{Qnn}\cdot \alpha ^{'}_{QQQ} & = & \left( \alpha ^{'}_{QQn}\right)
^2,
\eqry
\cite{Kos}, and \cite{BGH2} 
\bqry
\frac{1}{\alpha ^{'}_{nnn}}\;+\;\frac{1}{\alpha ^{'}_{QQn}} & = & \frac{2}{
\alpha ^{'}_{Qnn}}, \\
\frac{1}{\alpha ^{'}_{Qnn}}\;+\;\frac{1}{\alpha ^{'}_{QQQ}} & = & \frac{2}{
\alpha ^{'}_{QQn}}.
\eqry
We shall now show that only these counterparts of Eq. (5) hold in the heavy 
quark limit for baryons, not those of Eq. (4). 

Consider first Eqs. (24),(28). In the heavy quark limit $M(Q)\gg M(n),$ by 
virtue of the arguments given above in the meson case, it is possible to 
neglect the terms $\alpha ^{'}_{nnn}m^2_{nnn}$ and $1/\alpha ^{'}_{nnn}$ in 
Eqs. (24) and (28), respectively, in comparison with the remaining terms. 
Therefore, these equations reduce in this limit, respectively, to
\beq
\alpha ^{'}_{QQn}m^2_{QQn}\simeq 2\alpha ^{'}_{Qnn}m^2_{Qnn},
\eeq
\beq
\alpha ^{'}_{Qnn}\simeq 2\alpha ^{'}_{QQn},
\eeq
so that, independent of the values of the slopes,
\beq
m_{QQn}\simeq 2m_{Qnn}.
\eeq
Consider now Eqs. (25),(29). By expressing $\alpha ^{'}_{nnQ}m^2_{nnQ}$ and
$1/\alpha ^{'}_{nnQ}$ from Eqs. (24) and (28), and using further in (25),(29), 
one obtains, respectively,
\beq
\frac{1}{2}\;\!\alpha ^{'}_{nnn}m^2_{nnn}+\alpha ^{'}_{QQQ}m^2_{QQQ}=
\frac{3}{2}\;\!\alpha ^{'}_{QQn}m^2_{QQn},
\eeq
\beq
\frac{1}{2\alpha ^{'}_{nnn}}+\frac{1}{\alpha ^{'}_{QQQ}}=\frac{3}{2\alpha ^{
'}_{QQn}}.
\eeq
Neglecting again the terms containing $\alpha ^{'}_{nnn}$ in these relations, 
one obtains, independent of the values of the slopes,
\beq
m_{QQQ}\simeq \frac{3}{2}\;\!m_{QQn}.
\eeq
Eqs. (32),(35) imply
\beq
m_{QQQ}\simeq \frac{3}{2}\;\!m_{QQn}\simeq 3\;\!m_{Qnn},
\eeq
in agreement with the heavy quark limit $M(Q)\gg M(n).$ 

Now we apply a similar procedure to Eqs. (24)-(27). First, analogously to the
meson case, one obtains from (24),(26) in the heavy quark limit for baryons: 
\beq
\alpha ^{'}_{QQn}=\frac{\alpha ^{'}_{nnn}}{4},\;\;\;
\alpha ^{'}_{Qnn}=\frac{\alpha ^{'}_{nnn}}{2}.
\eeq
The use of these values of the slopes in Eq. (27) leads to
\beq
\alpha ^{'}_{QQQ}=\frac{\alpha ^{'}_{nnn}}{8}
\eeq
in this limit. Further use of $\alpha ^{'}_{QQn}$ and $\alpha ^{'}_{QQQ}$ in 
Eq. (33), which is also valid in this case, and in which we again neglect the 
term $\alpha ^{'}_{nnn}m^2_{nnn},$ results in
\beq
m_{QQQ}\simeq \sqrt{3}\;\!m_{QQn},
\eeq
in contradiction with the heavy quark limit, as given by (35). Thus, the four
equations (24)-(27) do not hold in the heavy quark limit for baryons, and 
therefore, only additivity of inverse Regge slopes, (5) and (28),(29), is 
consistent with the heavy quark limit for both mesons and baryons.

By using a parametrization of the baryon masses in terms of the current quark
mass which is similar to (8) \cite{BH}, and repeating the arguments given above
in the meson case, one can confirm that the equations (24),(25),(28),(29) also
hold in the formal chiral limit $m(n)\rightarrow 0.$
  
\section*{Concluding remarks}
We close with a brief summary of our results. 

We have shown that only additivity of inverse Regge slopes, (5) and (28),(29),
is consistent with the formal chiral, $m(n)\rightarrow 0,$ and heavy quark, 
$M(Q)>M(n),$ limits for both mesons and baryons. Alternative relations among 
the slopes, (4),(7) and their baryon counterparts, although consistent in the 
formal chiral limit, fail in the heavy quark limit.

We note, however, that empirical evidence for unambiguous preference of 
additivity of the inverse slopes is weak at present. Indeed, in the heavy quark
limit for mesons, Eq. (23) gives $\alpha ^{'}_{Q\bar{Q}}=\alpha ^{'}/4,$ and 
the slope of $b\bar{b}$ trajectory, as seen in Table I, is $\alpha ^{'}_{b\bar{
b}}=\alpha ^{'}/4.32.$ Eq. (22) in the same limit gives $m_{Q\bar{Q}}\simeq 2^{
5/4}\;\!m_{Q\bar{n}}\approx 2.378\;\!m_{Q\bar{n}},$ in contrast to $m_{Q\bar{
Q}}\simeq 2m_{Q\bar{n}}$ in this limit. Also, in the heavy quark limit for 
baryons, Eq. (39) gives $m_{QQQ}\simeq \sqrt{3}\;\!m_{QQn}\approx 1.732\;\!m_{
QQn},$ in contrast to $m_{QQQ}\simeq 1.5\;\!m_{QQn}$ in this limit. In any of  
the above three cases, the relative error does not exceed $\sim 15$\% (this is
the same accuracy as for mass relations derived on the basis of Eqs. (3),(4)
\cite{BGH1}). This situation is quite similar to the case of replacing an 
ultrarelativistic theory with $\langle {\bf p}^2\rangle /m^2\gg 1$ by a medium 
relativistic one with $\langle {\bf p}^2\rangle /M^2\approx 1,$ while the 
following remains valid: Even the lowest-order $1/M^2$ expansion is still 
legitimate with accuracy of the same order as above, $(1+\langle {\bf p}^2
\rangle /M^2)^{1/2}\approx 1+\langle {\bf p}^2\rangle /(2M^2),$ or $1.41\approx
1.5,$ with $\langle {\bf p}^2\rangle \sim M.$ This means an error of $\sim 6$\%
in the total energy and $\sim 20$\% in the kinetic energy; not excellent but 
sufficient for many purposes in hadronic physics, and, in particular 
\cite{LSG}, for hadron spectroscopy. That is why even alternative relations 
for Regge slopes lead to predictions for hadron masses which do not badly 
disagree with experiment, especially in the light quark sector where the forms
of these relations are almost indistinguishable from each other.


\bigskip
\bigskip
\begin{thebibliography}{9}
\bibitem{BB} J.-L. Basdevant and S. Boukraa, Z. Phys. C {\bf 28} (1985) 413; 
Ann. Phys. (Paris) {\bf 10} (1985) 475
\bibitem{JN} K. Johnson and C. Nohl, Phys. Rev. D {\bf 19} (1979) 291
\bibitem{KS} J.S. Kang and H. Schnitzer, Phys. Rev. D {\bf 12} (1975) 841
\bibitem{QR} C. Quigg and J.L. Rosner, Phys. Rep. {\bf 56} (1979) 167
\bibitem{rec} K.P. Das and R.C. Hwa, Phys. Lett. B {\bf 68} (1977) 459 \\
T. Tashiro {\it et al.,} Z. Phys. C {\bf 35} (1987) 21 \\
A.B. Batunin, B.B. Kiselyev and A.K. Likhoded, Yad. Phys. {\bf 49} (1989) 554
\bibitem{fra} A. Capella and J. van Tran Thanh, Z. Phys. C {\bf 10} (1981) 
249; Phys. Lett. B {\bf 114} (1982) 450 \\
A.B. Kaidalov, Phys. Lett. B {\bf 116} (1982) 459 \\
A.B. Kaidalov and K.A. Ter-Martirosyan, Sov. J. Nucl. Phys. {\bf 39} (1984) 
979 \\ A.B. Kaidalov and O.I. Piskounova, Z. Phys. C {\bf 30} (1986) 145 \\
G.I. Lykasov and M.N. Sergeenko, Z. Phys. C {\bf 52} (1991) 635
\bibitem{BGH1} L. Burakovsky, T. Goldman and L.P. Horwitz, Phys. Rev. D 
{\bf 56} (1997) 7119
\bibitem{BGH2} L. Burakovsky, T. Goldman and L.P. Horwitz, Phys. Rev. D
{\bf 56} (1997) 7124
\bibitem{BGH3} L. Burakovsky, T. Goldman and L.P. Horwitz, New Mass Relation
for Meson 25-plet [hep-ph/9704432], J. Phys. G, {\it in press}
\bibitem{BGH4} L. Burakovsky, T. Goldman and L.P. Horwitz, New Quadratic Mass
Relations for Heavy Mesons, FERMILAB-PUB-97-265-T [hep-ph/9708468]
\bibitem{KKY} K. Kawarabayashi, S. Kitakado and H. Yabuki, Phys. Lett. B {\bf
28} (1969) 432 
\bibitem{BESW} R.C. Brower, J. Ellis, M.G. Schmidt and J.H. Weis, Nucl. Phys. 
B {\bf 128} (1977) 175
\bibitem{KMP} N.A. Kobylinsky, E.S. Martynov and A.B. Prognimak, Ukr. Phys. 
Zh. {\bf 24} (1979) 969
\bibitem{DB} V.V. Dixit and L.A. Balazs, Phys. Rev. D {\bf 20} (1979) 816
\bibitem{IY} K. Igi and S. Yazaki, Phys. Lett. B {\bf 71} (1977) 158
\bibitem{Kai} A.B. Kaidalov, Z. Phys. C {\bf 12} (1982) 63
\bibitem{first} J. Pasupathy, Phys. Rev. Lett. {\bf 37} (1976) 1336 \\ 
K. Igi, Phys. Lett. B {\bf 66} (1977) 276; Phys. Rev. D {\bf 16} (1977) 196 
\bibitem{KY} M. Kuroda and B.-L. Young, Phys. Rev. D {\bf 16} (1977) 204
\bibitem{FS} S. Filipponi and Y. Srivastava, Hadronic Masses and Regge 
Trajectories, HUTP-97/A093 [hep-ph/9712204]
\bibitem{Scad} M.D. Scadron, Phys. Rev. D {\bf 29} (1984) 2076
\bibitem{BH} L. Burakovsky and L.P. Horwitz, Nucl. Phys. A {\bf 614} (1997)
373; On the Thermodynamics of Chiral Symmetry Restoration, LA-UR-96-2461
[hep-ph/9608290] 
\bibitem{prep} L. Burakovsky and T. Goldman, in preparation
\bibitem{Kos} R.S. Tutik, Ukr. Fiz. Zh. {\bf 26} (1981) 1937 \\
A.I. Kosenko and R.S. Tutik, Ukr. Fiz. Zh. {\bf 35} (1990) 1292
\bibitem{LSG} W. Lucha, F.F. Sch\"{o}berl and D. Gromes, Phys. Rep. {\bf 200}
(1991) 127
\end{thebibliography}
\end{document}